\documentclass[12pt]{article}
\usepackage{setspace}
\usepackage[margin=1.0in]{geometry}

\usepackage{siunitx}
	\DeclareSIUnit\molar{M}
	\DeclareSIUnit\torr{torr}
	\DeclareSIUnit\year{yr}
	\DeclareSIUnit\MWh{MWh}
	\DeclareSIUnit\GWh{GWh}
	\DeclareSIUnit\TWh{TWh}
	\DeclareSIUnit\MWp{MW\textsubscript{p}}
	\DeclareSIUnit\GWp{GW\textsubscript{p}}
	\DeclareSIUnit\Wac{W\textsubscript{ac}}
	\DeclareSIUnit\kWac{kW\textsubscript{ac}}
	\DeclareSIUnit\MWac{MW\textsubscript{ac}}
	\DeclareSIUnit\GWac{GW\textsubscript{ac}}
	\DeclareSIUnit\MMBtu{MMBtu}
	\DeclareSIUnit\ton{ton}
\usepackage[hyphens]{url}
\usepackage{hyperref}
	\hypersetup{colorlinks=false,pdfborder=0 0 0}
\usepackage{amsmath}
\usepackage{amsfonts,amsthm}
\usepackage{bm}
\usepackage{cleveref}
	\crefname{equation}{}{}
	\crefname{figure}{Figure}{Figures}	
	\crefname{table}{Table}{Tables}
	\crefname{section}{Section}{Sections}
	\crefname{chapter}{Chapter}{Chapters}
	\crefname{appendix}{SI~Note}{SI~Notes}
\usepackage[caption=false,font=normalsize,labelfont=sf,textfont=sf]{subfig}
\usepackage{graphicx}
\usepackage[version=4]{mhchem}
\usepackage{textcomp}
\usepackage{xfrac}
\usepackage{microtype}
\usepackage{booktabs}
\usepackage{xcolor}
\usepackage{authblk}
\newcommand{\about}{\raise.17ex\hbox{$\scriptstyle\sim$}}
\usepackage{mathtools}

\title{A general method for estimating zonal transmission interface limits from nodal network data}

\author{Patrick R. Brown\thanks{patrick.brown@nrel.gov}}
\author{Clayton P. Barrows\thanks{clayton.barrows@nrel.gov}}
\author{Jarrad G. Wright}
\author{\mbox{Gregory L. Brinkman}}
\author{Sourabh Dalvi}
\author{Jiazi Zhang}
\author{Trieu Mai}
\affil{National Renewable Energy Laboratory}
\date{}

\begin{document}

\maketitle
\begin{abstract}
\setstretch{1.3}
Capacity expansion models for the electric power system often employ zonal (rather than nodal) resolution, necessitating estimates of aggregate power transfer limits across the interfaces between model zones.
Interface limits between planning areas are sometimes published, but they are not generalizable to arbitrary zone shapes.
There is thus a need for a reproducible method for estimating interface transfer limits (ITLs) between user-defined zones directly from nodal transmission system data.
Here, we present a simple method for estimating ITLs using a DC power flow approximation via the power transfer distribution factor (PTDF) matrix.
Linear optimization is performed to identify the distribution of power flows that maximizes the total flow on interface-crossing lines, subject to individual line ratings, limits on bus injection/withdrawal, and the relationships among flows, injections, and withdrawals imposed by the PTDF matrix.
We demonstrate the application of the method on a 134-zone \about\num{65000}-bus system, and we explore the influence of flow direction, contingency level, and zone size on the estimated ITLs.
There is significant heterogeneity in the ratio of the ITL to the sum of interface-crossing line ratings, which highlights the importance of accounting for the physical constraints on power flows imposed by Kirchhoff's laws when estimating zonal ITLs.
\end{abstract}
\setstretch{1.8}


\section*{Nomenclature} \label{sec:nomenclature}
\subsection*{Indices and sets}

\begin{tabular}{p{0.22\linewidth}p{0.6\linewidth}}
    $z \in Z$ & Zones \\
    $i \in I$ & Interfaces, where $i_{z',z''}$ is the interface between zones $z'$ and $z''$ (sometimes indicated by $z'||z''$) \\
    $b \in B$ & Buses \\
    $b_v \in B_v \subset B$ & Generator buses \\
    $b_q \in B_q \subset B$ & Load buses \\
    $b_t \in B_t \subset B$ & Transmission buses \\
    $l \in L$ & Transmission lines and transformers (both typically referred to as ``lines''), where $l^{b',b''}$ is a line defined from starting bus $b'$ to ending bus $b''$ \\
    $l_{i} \in L_i \subset L$ & Lines crossing interface $i_{z',z''}$
\end{tabular}

\subsection*{Parameters}
\begin{tabular}{p{0.22\linewidth}p{0.6\linewidth}}
    $p(l,b)$ & PTDF matrix value for line $l$ and bus $b$ \\
    $r(l)$ & Rating for line $l$ [\si{\mega\watt}]
\end{tabular}

\subsection*{Variables}
\begin{tabular}{p{0.22\linewidth}p{0.6\linewidth}}
    $F(l)$ & Power flow on line $l$ [\si{\mega\watt}] (positive when power flows from $b'$ to $b''$ on line $l^{b',b''}$; negative when power flows from $b''$ to $b'$) \\
    $G(b)$ & Net injection at bus $b$ [\si{\mega\watt}] (positive for generation, negative for load)
\end{tabular}

\section{Introduction} \label{sec:intro}

Multidecadal capacity expansion models (CEMs) for interregional power system planning typically must sacrifice spatial resolution to achieve tolerable problem sizes and solution times \cite{Ho2021,Young2020,EPAIPM2021}.
Rather than model individual generators, loads, and transmission lines, modelers group power systems into geographic areas (here referred to as ``zones'').
The zones themselves are treated as ``copper plates'', without constraints on intrazonal power flows;
power flows across zonal interfaces are constrained by interface transfer limits (here abbreviated as ITLs), which are meant to approximate the combined carrying capacity of interface-crossing transmission lines under standard operating conditions.

The selection of model zone boundaries is often constrained by the availability of underlying input data;
hourly load profiles may only be available for balancing areas or independent system operators (ISOs),
emissions policies may apply within state boundaries,
and installed generation capacity may be available by county or state.
ITLs may be published for individual transmission corridors, as in the case of intertie paths in the Western Electricity Coordinating Council (WECC) \cite{WECCpath2022},
or for the interfaces between balancing areas or regional transmission organizations (RTOs), but they are not generalizable to the zone shapes that may be needed to represent other model features.
Policymakers are also interested in understanding ITLs between different planning regions and their impact on resource adequacy and reliability \cite{FERCworkshop2022}.
Zonal ITLs necessarily represent a simplification compared to nodal power flow and should be treated as approximate estimates rather than precise limits applicable under all operating conditions; nevertheless, their utility across a range of models necessitates a reproducible estimation method.

In general, all transmission lines crossing an interface may not necessarily be used at their rated capacities simultaneously.
In large meshed networks, the physics of electrical power flow tend to result in some lines reaching their rated capacity before others, thus preventing the others from contributing their full rated capacity to the ITL.
Maximum power flows may also be direction-dependent; if an outlying zone has only load buses (or generator buses), it may only be able to accept power from (or send power to) connected zones.
Thus, ITLs often cannot be accurately approximated by the sum of steady-state ratings of all transmission lines crossing an interface.

The convenience of quantifying interzonal transfer limits has been recognized in planning and operations procedures for decades, and there is an associated rich body of literature on ITL definitions \cite{ilic_transmission_1998} and calculation methods \cite{austria_integrated_1995, gan_stability-constrained_2000, gan_minmax_2003, mohammed_available_2019}.
A comprehensive review \cite{mohammed_available_2019} compares ITL calculation methods and the generation, load, and transmission details considered by each one.
However, the discussed methods focus on parametric evaluation of transfer capability based on incrementally increasing power injected and withdrawn at corresponding generation and load buses.
Exceptions are presented in \cite{gan_stability-constrained_2000, gan_minmax_2003}, where maximum flow problems are formulated to calculate the ITL.
Nonetheless, the maximum flow problems are formulated for individual interfaces between bus or zone pairs and little discussion is devoted to the scalability and generality of the presented methods.
ITLs are used as the reference from which available transfer capability (ATC) is calculated as the ITL minus the sum of the transfer reserve margin and existing transmission commitments.
The prevalent use of ATC in electric industry planning and operations procedures \cite{noauthor_pjm_nodate} suggests accurate ITL calculation is critical.
Furthermore, in an industry that increasingly relies on a distributed set of resources and is subject to a range of policies applied over a variety of geographic regions, ITL calculation should be both scalable and general so that ITLs can be frequently updated and are not overly sensitive to system dispatch conditions.
Thus, there is a need for a reproducible method for estimating ITLs, obeying the physical constraints on power flows imposed by Kirchhoff's laws.

Here, we present a simple method for calculating ITLs between arbitrarily-defined zones, combining individual line ratings, bus types, and a representation of DC power flow by way of the power transfer distribution factor (PTDF) matrix derived from the individual branch impedances and network topology (\cref{sec:itl}).
The method produces ITLs for both power flow directions across a given interface, and it can be used at the $n-0$, $n-1$, or higher contingency levels.
We demonstrate its use on a simple 5-bus 3-zone test system \cite{Li2010} (\cref{sec:results-5bus}), and on the \about\num{65000}-bus 2024 system used in the North American Renewable Integration Study (NARIS) \cite{Brinkman2021} (\cref{sec:data}) combined with the 134 default zones of the Regional Energy Deployment System (ReEDS) model \cite{Ho2021} (\cref{sec:results-naris}).
Computer code for implementing the PTDF-based method is made available open source \cite{Barrows2023}, and it can be used on both empirical (typically proprietary) network data and on open-source synthetic test systems such as the ``TAMU'' dataset \cite{TAMU,Xu2020}.

\section{Methods and input data} \label{sec:methods}

\subsection{Calculation of interface transfer limits} \label{sec:itl}

\subsubsection{Conventions} \label{sec:conventions}

A line segment is said to cross interface $i_{z',z''}$ if one of its endpoints is in zone $z'$ and its other endpoint is in zone $z''$.
Each line $l^{b',b''}$ has a starting bus $b'$ and an ending bus $b''$; flow on line $l^{b',b''}$ is positive if power flows from bus $b'$ to bus $b''$ and negative if power flows from $b''$ to $b'$.
(We refer to all network branches as lines, even though some branches represent transformers.)
For most lines, we use the polarity of starting and ending buses defined by the underlying nodal dataset,
but interface-crossing lines are always redefined such that the starting bus is in the first alphanumerically sorted zone and the ending bus is in the second alphanumerically sorted zone.

\subsubsection{Objective function} \label{sec:objective}

For each interface $i_{z',z''}$ we perform independent optimizations to determine the ITL in the ``forward'' direction (from $z'$ to $z''$) and ``reverse'' direction (from $z''$ to $z'$).
In the forward direction the objective is to maximize the sum of flows on lines crossing interface $i_{z',z''}$:
\begin{subequations} \label{eq:objective}
    \begin{align}
        \max_{F,G} \sum_{l_i}^{L_{i}} F(l_i) \\
        \intertext{In the reverse direction the objective is to minimize the sum of flows on lines crossing interface $i_{z',z''}$, recalling that negative flows represent flows from $z''$ to $z'$:}
        \min_{F,G} \sum_{l_i}^{L_{i}} F(l_i)
    \end{align}
\end{subequations}
These optimizations are performed independently for each interface $i$ in the set of all interfaces $I$.

\subsubsection{Constraints} \label{sec:constraints}
Line ratings are enforced for all lines included in the simulation.
Power can flow along an individual line in either direction, subject to its rating:
\begin{align} \label{eq:rating}
    -r(l) & \leq F(l) \leq r(l) & \forall l
\end{align}

The relationship between line flows and bus injection/withdrawal is governed by the PTDF matrix:
\begin{align} \label{eq:ptdf}
    F(l) & = \sum_{b}^B p(l,b) G(b) & \forall l
\end{align}
The PTDF matrix provides a simple linear representation of DC power flow \cite{Christie2000}; calculated from the reactance matrix as described in textbooks (e.g. \cite{Wood2014}), it relates the flows on all lines $l$ to the injections/withdrawals on all buses $b$ in an AC network.
The PTDF matrix element $p(l',b')$ gives the change in power flow along line $l'$ that accompanies a 1-unit change in power injection from bus $b'$.

As we are interested in the physical capacity of the transmission network itself rather than typical operating conditions, the level of power injection/withdrawal at each bus is not constrained by the installed generation capacity or load participation factor.
But whenever data on the bus type is available, the bus is constrained to either inject power (for generator buses), withdraw power (for load buses), or to neither inject nor withdraw (for transmission buses):
\begin{subequations} \label{eq:bus}
    \begin{align}
        G(b_v) & \geq 0 & \forall b_v \in B_v \\
        G(b_q) & \leq 0 & \forall b_q \in B_q \\
        G(b_t) & = 0 & \forall b_t \in B_t
    \end{align}
\end{subequations}
If no information on bus type is available for a given bus, or if a bus qualifies as both a generator and load bus, constraint \cref{eq:bus} is not applied for that bus.

Together, constraints \cref{eq:ptdf,eq:bus} tend to limit the ITL to a value less than the simple sum of interface-crossing line ratings.

\subsection{Example using 5-bus test system} \label{sec:results-5bus}

We first illustrate the method for a simple 5-bus test system \cite{Li2010}, with topology and line parameters (reactance and power rating) defined in \textbf{\cref{tab:5bus-inputs}}.
\begin{table}
    \centering
    \caption{5-bus test system line parameters.}
      \begin{tabular}{l|p{1.0cm}p{1.0cm}p{1.0cm}p{1.0cm}p{1.8cm}p{1.0cm}}
      \toprule
      Line  & From bus & To bus & From zone & To zone & Reactance & Rating [MW] \\
      \midrule
      A$|$B   & A     & B     & 1     & 1     & 0.0281 & 400 \\
      B$|$C   & B     & C     & 1     & 1     & 0.0108 & 400 \\
      C$|$D   & C     & D     & 1     & 2     & 0.0297 & 400 \\
      D$|$E   & D     & E     & 2     & 3     & 0.0297 & 240 \\
      A$|$E   & A     & E     & 1     & 3     & 0.0064 & 400 \\
      A$|$D   & A     & D     & 1     & 2     & 0.0304 & 400 \\
      \bottomrule
      \end{tabular}%
    \label{tab:5bus-inputs}%
  \end{table}%
The corresponding PTDF matrix for this network is provided in \textbf{\cref{tab:5bus-ptdf}}, using bus A as the slack bus (the results are independent of the choice of slack bus).
This example does not include limitations on injection/withdrawal by bus type; each bus is allowed to act as either generation or load, so constraint \cref{eq:bus} is inactive and the ITL for each interface is the same in both directions.
\begin{table}
  \centering
  \caption{PTDF matrix for the 5-bus test system.}
    \begin{tabular}{l|rrrrr}
      \toprule
          & \multicolumn{1}{l}{A} & \multicolumn{1}{l}{B} & \multicolumn{1}{l}{C} & \multicolumn{1}{l}{D} & \multicolumn{1}{l}{E} \\
      \midrule
    A$|$B   & 0     & -0.6698 & -0.5429 & -0.1939 & -0.0344 \\
    B$|$C   & 0     & 0.3302 & -0.5429 & -0.1939 & -0.0344 \\
    C$|$D   & 0     & 0.3302 & 0.4571 & -0.1939 & -0.0344 \\
    D$|$E   & 0     & 0.1509 & 0.2090 & 0.3685 & -0.1120 \\
    A$|$E   & 0     & -0.1509 & -0.2090 & -0.3685 & -0.8880 \\
    A$|$D   & 0     & -0.1792 & -0.2481 & -0.4376 & -0.0776 \\
    \bottomrule
    \end{tabular}%
  \label{tab:5bus-ptdf}%
\end{table}%

\begin{figure}[tbp]
   \centering
   \includegraphics[width=0.7\linewidth]{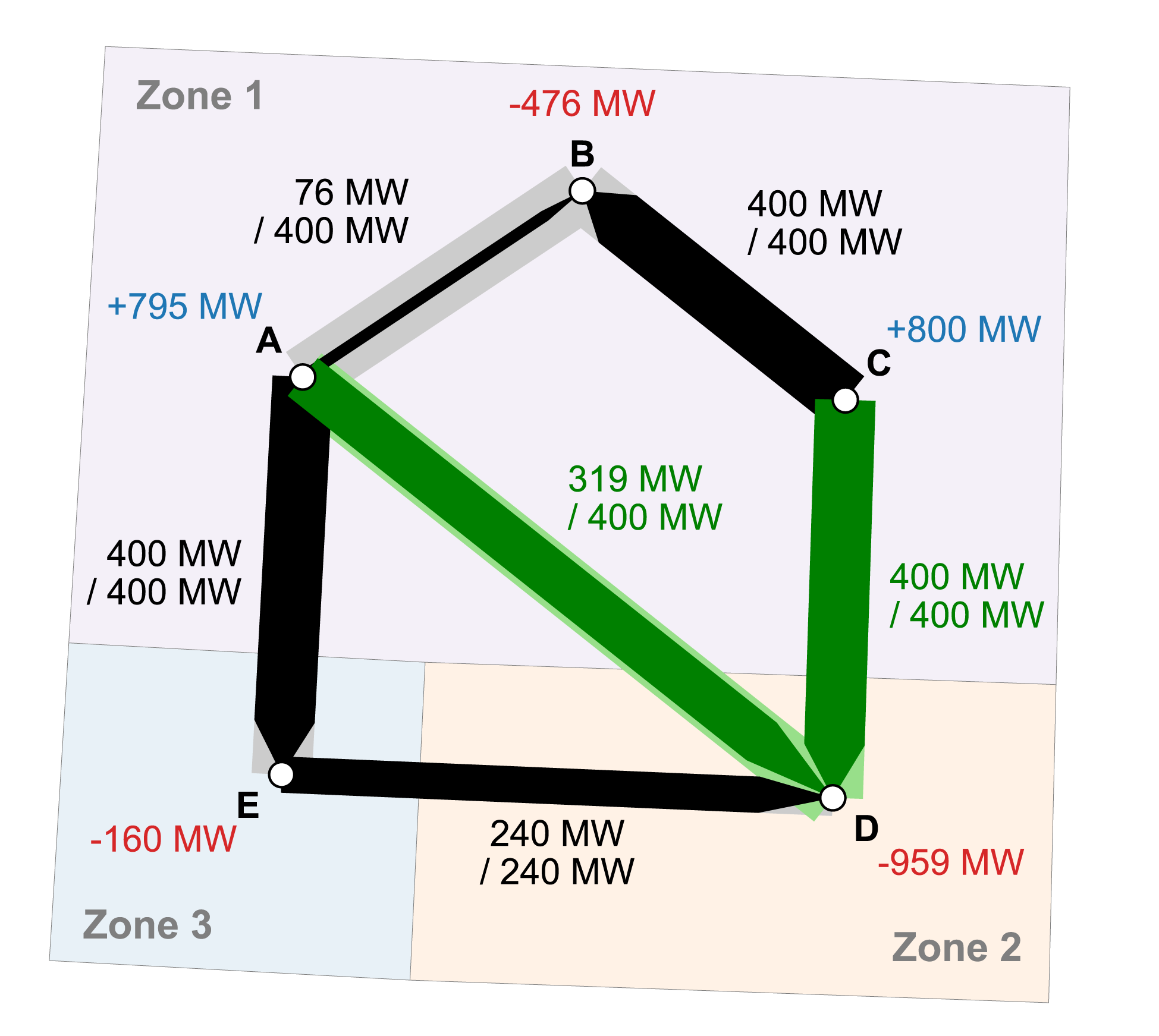}
   \caption{
       \textbf{Line flows in the 5-bus test system corresponding with the ITL of the $\bm{1||2}$ interface.}
       The first number in each black or green pair gives the flow on the indicated line; the second number gives the line rating.
       Lines crossing the $1||2$ interface are shown in green; lines not crossing the $1||2$ interface are shown in black.
       The width of the gray and light green bars indicates the line rating; the width and color of the overlaid arrow indicate the flow from source (wide end of arrow) to sink (narrow end of arrow).
       Blue numbers indicate power injection (+) and red numbers indicate power withdrawal (–) at each bus.
   }
   \label{fig:f_map-5bus-flows}
\end{figure}

The resulting ITLs are given in \textbf{\cref{tab:5bus-itl}}.
Each ITL is associated with an independent set of line flows, with flows corresponding to the ITL of the $1||2$ interface shown in \textbf{\cref{fig:f_map-5bus-flows}}.
\begin{table}
    \centering
    \caption{5-bus test system transfer limits.}
      \begin{tabular}{l|p{2.2cm}p{1.8cm}p{1.2cm}}
        \toprule
        Interface  & Interface-crossing lines & \mbox{Sum of} \mbox{ratings} [MW] & ITL [MW] \\
        \midrule
        1$||$2 & A$|$D, C$|$D &800 & 719 \\
        2$||$3 & D$|$E & 240 & 400 \\
        1$||$3 & A$|$E & 400 & 240 \\
        \bottomrule
      \end{tabular}%
    \label{tab:5bus-itl}%
  \end{table}%
Note that while the total flow from Zone 1 to Zone 2 (including flow through Zone 3) is \SI{959}{\mega\watt}, the ITL only includes the contribution from interface-crossing lines A$|$D and C$|$D.
In this case, line C$|$D is fully utilized under interface-flow-maximizing conditions, but line A$|$D is underutilized, contributing \SI{319}{\mega\watt} of a potential \SI{400}{\mega\watt} to the ITL.
As a result of the constraints imposed by the PTDF matrix (i.e. the impact of Kirchhoff's voltage law on power flows), increasing the flow on line A$|$D would require increasing the flow on lines A$|$E (in the positive direction) and D$|$E (in the negative direction, from bus E to bus D), both of which are already at their rated limit.
Increasing the ITL between Zones 1 and 2 could thus be accomplished by increasing the rating of line C$|$D; changing the reactance of lines A$|$D, A$|$E, and D$|$E; or increasing the ratings of lines A$|$E and D$|$E; but not by increasing the rating of interface-crossing line A$|$D.

\subsection{Demonstration data for contiguous U.S.}
\label{sec:data}

\begin{figure}[tbp]
   \centering 
   \includegraphics[width=0.9\linewidth]{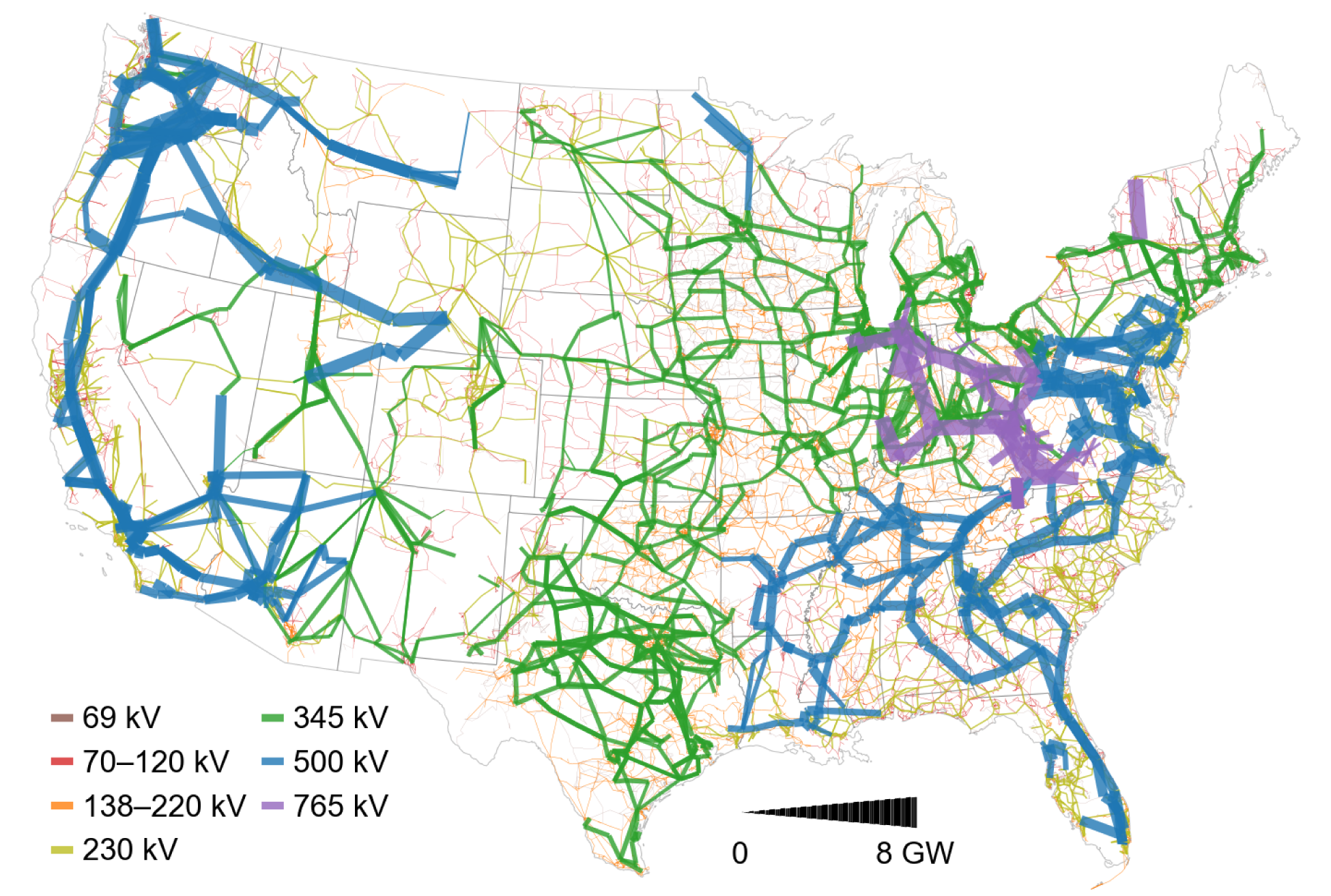}
   \caption{
       \textbf{Transmission line dataset for the contiguous U.S.}
       The rating of each transmission line is indicated by the line width; the voltage is indicated by the color.
       The data shown here were assembled as part of the North American Renewable Integration Study (NARIS) \cite{Brinkman2021}.
   }
   \label{fig:naris-lines}
\end{figure}

\begin{figure}[tbp]
   \centering 
   \includegraphics[width=0.8\linewidth]{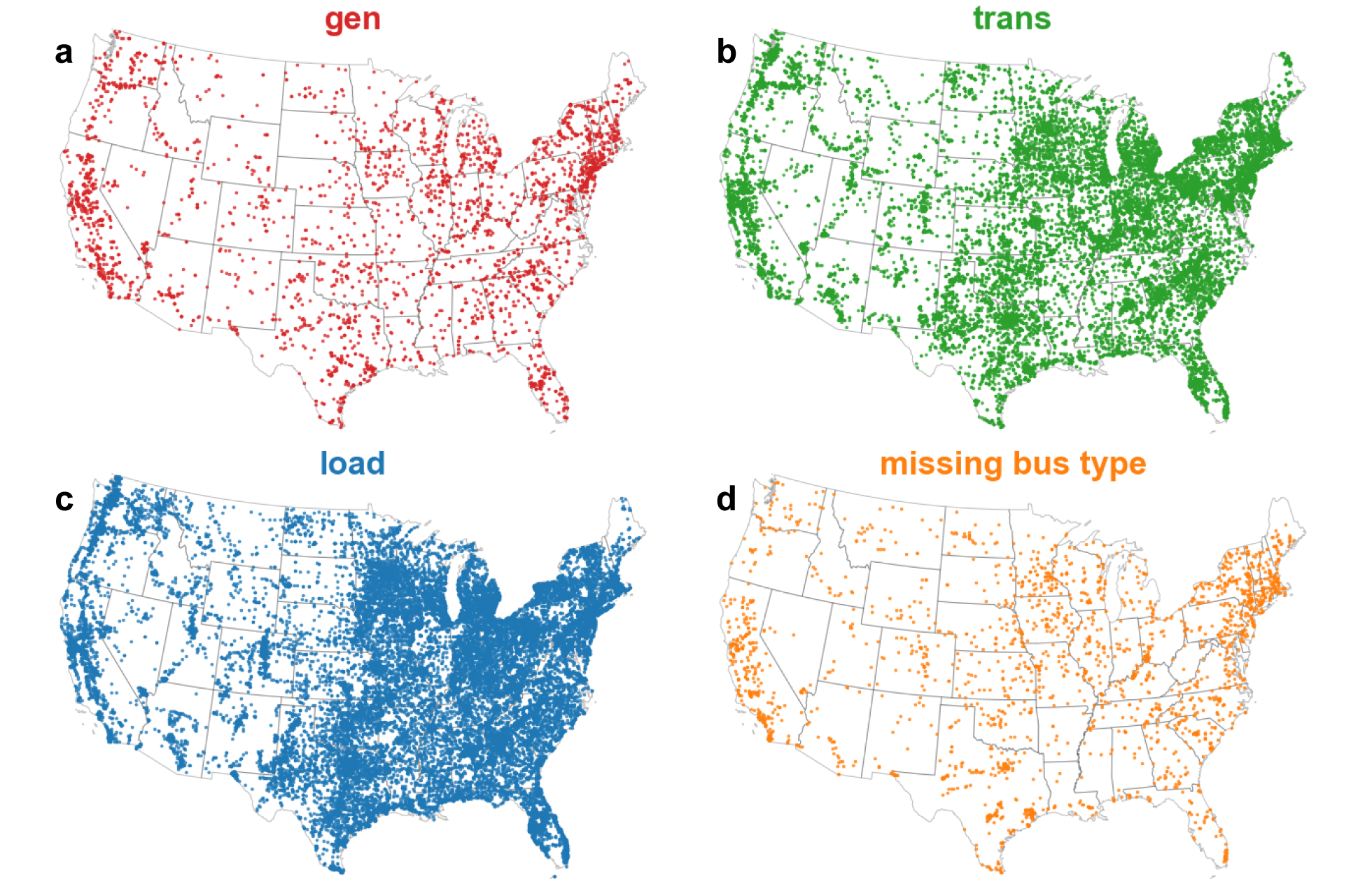}
   \caption{
       \textbf{Bus dataset for the contiguous U.S.} \cite{Brinkman2021}
   }
   \label{fig:naris-nodes}
\end{figure}

The lines and buses included in the NARIS nodal dataset \cite{Brinkman2021} are shown in \textbf{\cref{fig:naris-lines}} and \textbf{\cref{fig:naris-nodes}} respectively.
The following data-cleaning steps were employed:
\begin{itemize}
    \item Buses without location data were assigned to the location of the bus separated by the smallest number of ``hops'' through the network. If a bus without location data is connected to two or more buses by the same number of network hops (e.g. if bus A, without location data, is connected to buses B and C, which do have location data), the missing-location bus was assigned to a point located at the centroid of the connected buses with location data.
    \item Buses located more than \SI{100}{\kilo\meter} outside the contiguous U.S. were dropped from the dataset.
    \item Network reduction was performed to remove radial buses connected to only a single other bus, reducing problem size. For the purposes of constraint \cref{eq:bus}, if a generator bus was removed, the bus it was attached to is treated as a generator; if a load bus was removed, the bus it was attached to is treated as a load.
    \item DC lines and back-to-back interties were dropped, as the method described here only applies to AC transmission. For application in a CEM or production cost model (PCM), DC lines and interties would be modeled separately, with flow limits defined by their rated capacities.
    \item For lines without known rating data, missing ratings were inferred using practical capability limits for transmission lines based on the loadability concepts introduced by St. Clair \cite{StClair1953} (\cref{fig:f_loadability} in the Supplemental Information). Normalized line characteristics at typical subtransmission and transmission voltages were based on \cite{Hao2008} while calculations for thermal and surge impedance loading (SIL) ratings were based on \cite{Saadat1999}. Line lengths were calculated as the geodesic distance between the two line end points, likely underestimating the actual length and hence slightly overestimating actual ratings.
    \item For lines without reactance data, the missing reactance was copied from the line of the same voltage with the length nearest to the length of the line without reactance data. For transformers without reactance data (which have zero length), the missing reactance was taken as the median across transformers of the same voltage that include reactance data.
    \item The three interconnections (the Eastern Interconnection, Western Interconnection, and Texas Interconnection) were simulated independently. When assessing interfaces within the Eastern Interconnection, only buses within \SI{800}{\kilo\meter} of the specified interface were included in the simulation to reduce problem size. For interfaces within the Western Interconnection and Texas Interconnection, all buses within the interconnection were included.
\end{itemize}

\begin{figure}[tbp]
    \centering
    \includegraphics[width=0.8\linewidth]{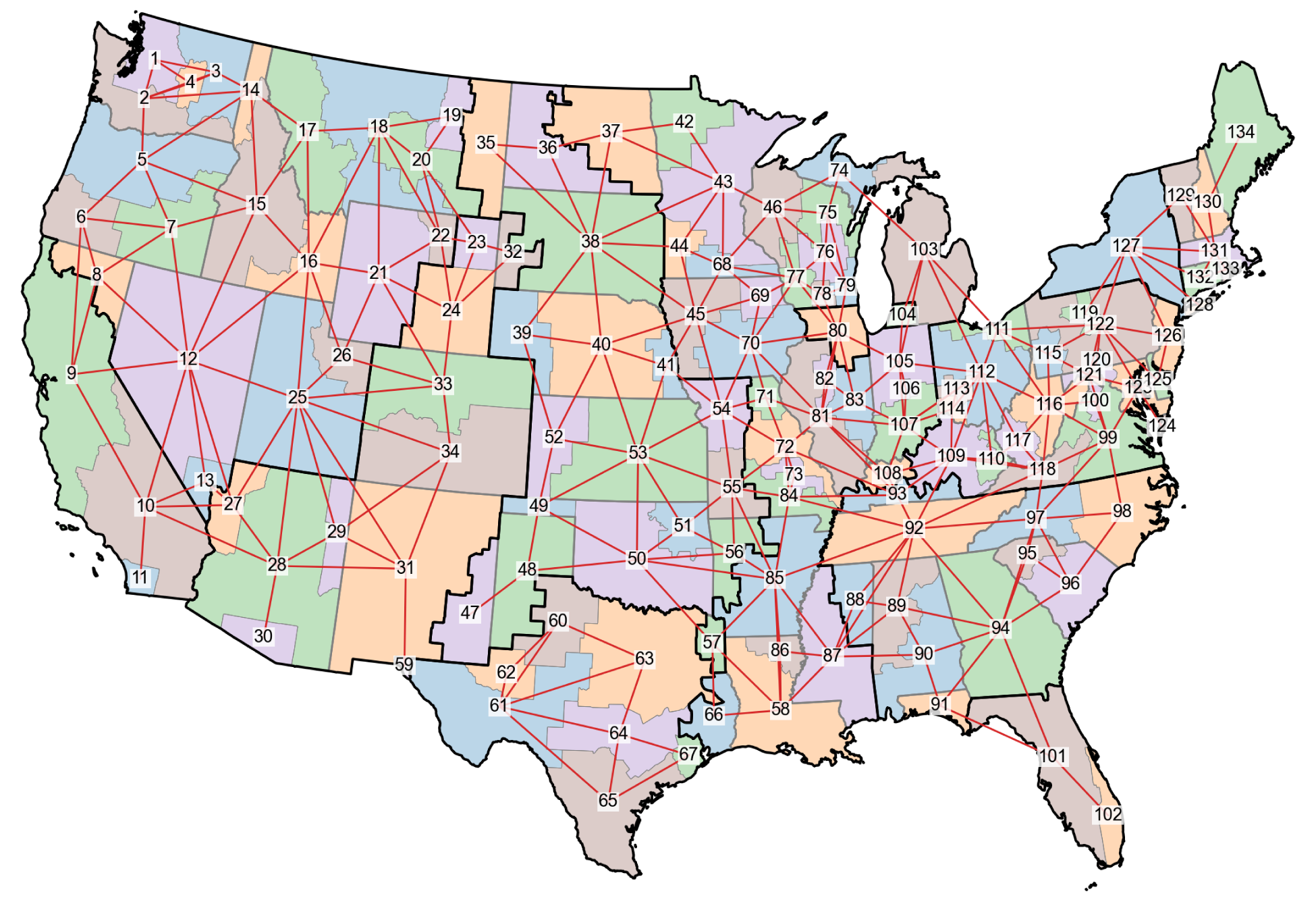}
    \caption{
        \textbf{Default model zones used in ReEDS.}
        Adjacent zones within the same interconnection are connected by red lines.
    }
    \label{fig:f_map-zones_reeds}
\end{figure}

The ITL calculation described here can be applied to arbitrary zone shapes ranging from political boundaries (counties or states) to actual balancing areas.
For demonstration purposes, we apply it to the 134 default zones used in the ReEDS model, shown in \textbf{\cref{fig:f_map-zones_reeds}} \cite{Ho2021}.
For each pair of zones within the same interconnection connected by at least one line, we first calculate the ITL in both directions (i.e. from the alphanumerically first zone to the second zone, and vice versa), including all lines in the network (henceforth referred to as ``$n-0$'' conditions).
For each interface and direction, we then identify the interface-crossing line with the highest flow, remove it from the network, and recalculate the ITL.
These (typically lower) ratings are referred to as ``$n-1$'' ratings.
Because intrazonal, non-interface-crossing lines could also be binding, a true contingency analysis would repeat the calculation for every line in the network, not just interface-crossing lines.
As this approach would increase computation times by many orders of magnitude, we do not employ it here; instead we focus our contingency analysis on interface-crossing lines alone.

\subsection{Caveats and limitations}
\label{sec:caveats}
Before discussing the resulting ITLs for the ReEDS interfaces, we first note some limitations of the method described here.
\begin{itemize}
    \item
    This method does not consider the use of phase-shifting transformers or other flexible AC transmission system (FACTS) devices that can modify power flow over individual branches or groups of branches. Increased use of FACTS devices would relax constraint \cref{eq:ptdf} and could increase ITLs.
    \item
    As we are concerned with the physical capabilities of the transmission network, we do not place limits on the level of injections/withdrawals to or from individual buses beyond constraint \cref{eq:bus}.
    For the existing system, injections would be limited by the amount of installed generation capacity (combined with weather conditions for variable renewable energy resources), and withdrawals would likely be distributed according to historically observed load participation factors rather than varying arbitrarily in order to enable maximum interface flows.
    Including limits based on installed generation capacity and observed withdrawals (load participation factors) would be expected to reduce the calculated ITLs.
    \item
    The method used to infer missing branch ratings (described in the Supplemental Information) considers the combination of individual branch thermal and angular stability limits but not voltage stability limits.
    This approach provides an upper limit on individual branch ratings as a function of line length.
    Branch ratings reported in the NARIS dataset are used as-is, assuming the rating includes applicable de-rating beyond thermal limits.
    It is expected that a majority of individual branch ratings would be based on thermal limits, with exceptions for long lines that could include appropriate de-ratings relative to thermal limits.
    If a reported branch rating only includes the thermal limit and neglects applicable de-rating, the resulting ITL could be overestimated.
    \item
    Zonal capacity-expansion or production-cost simulations using interzonal ITLs offer lower resolution than a full nodal model since zonal models treat the area within zones as congestion-free ``copper plates.''
    When possible, it is preferable to locate zone boundaries such that they intersect important congestion-limited corridors; otherwise, the ITL could give an inflated estimate of the actual ability to move power across a zone with the existing network.
    For example, if a zone were defined such that it encompassed the boundary between two asynchronous interconnections, the ability to flow power across that zone between interconnections would be limited not only by the ITLs on either side of the zone but also by the capacity of any back-to-back interties located within the zone.
    Thus, while the method described here can be used for any specified zone boundaries, the utility of the estimated ITL will depend on how well the specified zones intersect with congestion patterns of interest.
\end{itemize}

\section{Results} \label{sec:results}

\subsection{ITLs resolved by flow direction and contingency level}
\label{sec:results-naris}
\begin{figure*}[tbp]
   \centering
   \includegraphics[width=0.85\linewidth]{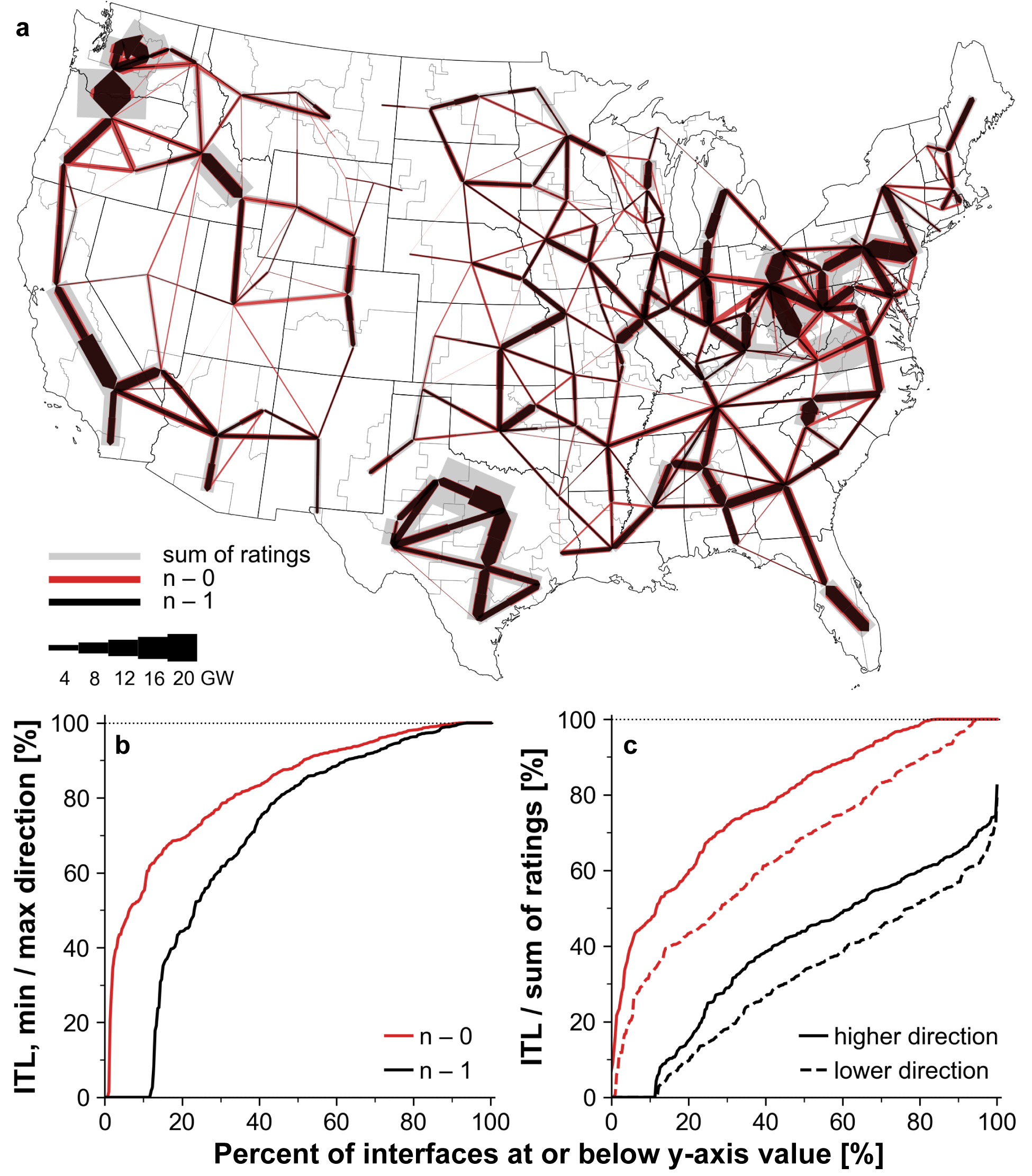}
   \caption{
       \textbf{ITLs for ReEDS zones as a function of direction and contingency level.}
       \textbf{a}, Map showing the sum of ratings for interface-crossing lines (gray), the ITL at $n-0$ contingency level (red), and the ITL at $n-1$ contingency level (black).
       Each interface is represented by a group of bars drawn between the centroids of the two zones it connects, with the thickness indicating the rating; ITLs are shown for each direction, with the ITL flowing into a zone indicated by the thickness of the half of the bar pointing into that zone.
       \textbf{b}, Distribution of the ratios of the ITL between the higher-limit and lower-limit directions for each interface, for both $n-0$ (red) and $n-1$ contingency levels (black).
       \textbf{c}, Distribution of the ratios between the ITL and the sum of ratings of interface-crossing lines for that interface.
       Red lines indicate $n-0$ conditions; black lines indicate $n-1$ conditions.
       Solid lines indicate ratios calculated using the higher-capacity direction; dashed lines indicate ratios calculated using the lower-capacity direction.
       The distribution on the $x$ axis includes 293 interfaces between the 134 ReEDS model zones that are connected by at least one AC transmission line.
   }
   \label{fig:f_mapdist-itl_reeds-contingency}
\end{figure*}

\textbf{\Cref{fig:f_mapdist-itl_reeds-contingency}} shows the difference between the sum of line ratings, the ITL at $n-0$ contingency level, and the ITL at $n-1$ contingency level for the 293 synchronous zonal interfaces associated with the 134 ReEDS model zones.

There is substantial heterogeneity in the impact of both the direction of flow and the contingency level on ITLs across the zonal interfaces assessed here.
Under $n-0$ conditions, \SI{31}{\percent} of interfaces have ITLs that vary between the two directions of flow by $\leq \SI{5}{\percent}$ ,
while \SI{6}{\percent} of interfaces have ITLs that vary between the two directions by more than a factor of 2 (red trace in \cref{fig:f_mapdist-itl_reeds-contingency}b).
In the higher-flow direction, \SI{16}{\percent} of interfaces have an ITL equal to the sum of ratings for interface-crossing lines,
while \SI{12}{\percent} of interfaces have an ITL equivalent to less than half of the sum of ratings for interface-crossing lines (red solid trace in \cref{fig:f_mapdist-itl_reeds-contingency}c).
The median ratio of the ITL to the sum of interface-crossing line ratings at the $n-0$ contingency level is \SI{84}{\percent} for the higher-ITL direction and \SI{69}{\percent} for the lower-ITL direction.

Under $n-1$ conditions, the ITL is typically substantially smaller than the sum of interface-crossing line ratings:
the median ratio is \SI{44}{\percent} for the higher-ITL direction and \SI{33}{\percent} for the lower-ITL direction.
\SI{12}{\percent} of interfaces have an ITL of zero in at least one of the two directions (black trace in \cref{fig:f_mapdist-itl_reeds-contingency}b).

\begin{figure}
   \centering
   \includegraphics[width=0.85\linewidth]{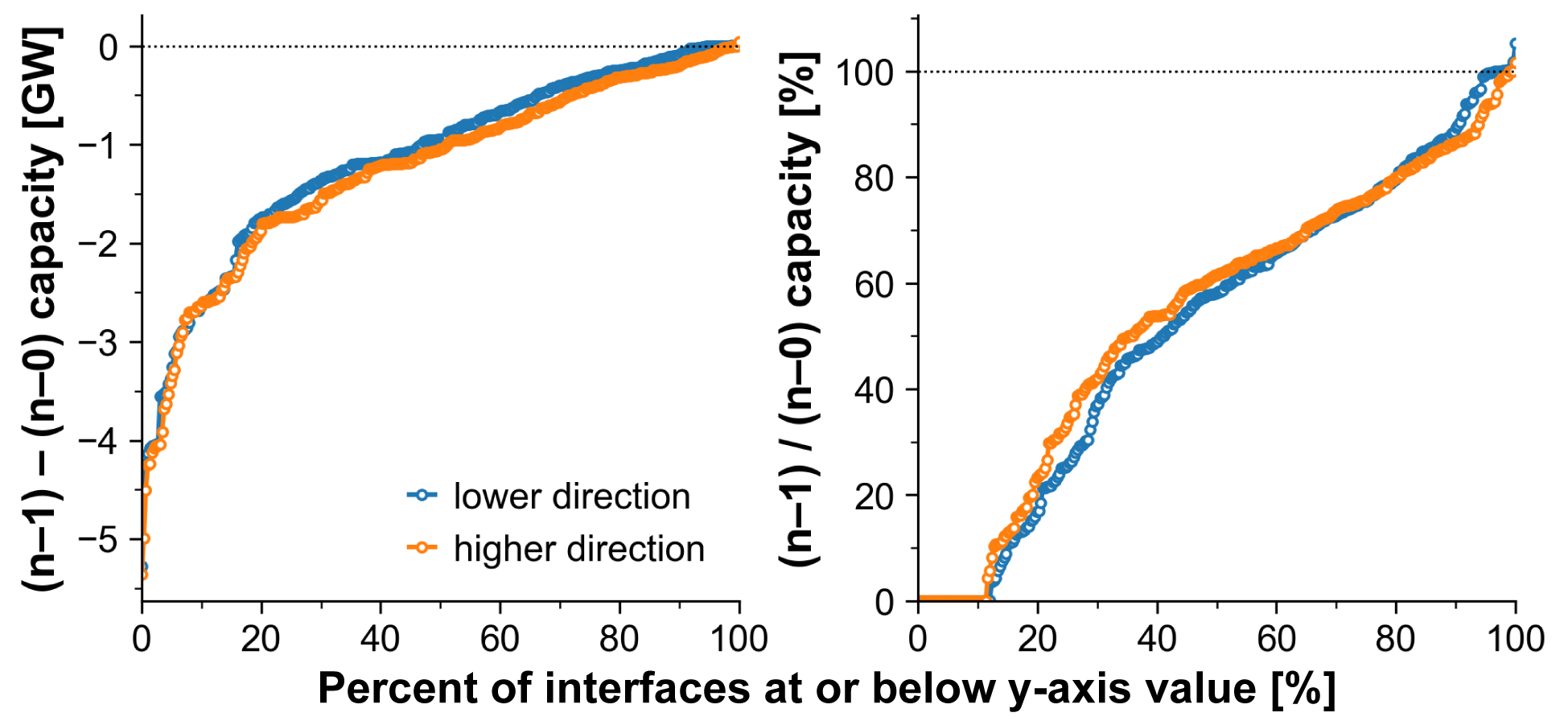}
   \caption{
       \textbf{Distribution of the impacts of subtracting the line with largest flow from each interface, shown for the direction with lower $\bm{n-0}$ rating (blue) and the direction with higher $\bm{n-0}$ rating (orange).}
       \textbf{a}, Difference between $n-1$ and $n-0$ ratings.
       \textbf{b}, Ratio between $n-1$ and $n-0$ ratings.
       The distribution includes 293 interfaces between the 134 ReEDS model zones that are connected by at least one AC transmission line.
   }
   \label{fig:f_dist-contingency-ratio}
\end{figure}

\textbf{\Cref{fig:f_dist-contingency-ratio}} shows the impact of applying $n-1$ contingency conditions relative to the $n-0$ ITL.
The median reduction in ITL for $n-1$ conditions is 0.9--\SI{1.1}{\giga\watt}, or 39--\SI{42}{\percent}, depending on the direction of flow.
As noted above, the reduction is much larger for a small collection of interfaces;
9--10 interfaces (depending on the flow direction) lose $\geq \SI{4}{\giga\watt}$ of ITL capacity at the $n-1$ contingency level.

\subsection{Aggregating transfer limits of smaller zones to obtain transfer limits for larger zones} \label{sec:results-zones}

Accurate ITLs are likely to be useful at a variety of geographic resolutions.
ITLs for smaller zones can be useful for assessing the trade-off between installing renewable energy capacity in high-quality resource regions far from load (incurring transmission expansion costs) and installing capacity in lower-quality resource regions nearer to load;
ITLs between larger planning regions (such as ISOs or RTOs) may be useful for longer-term interregional resource planning.
Having calculated the ITLs between the 134 ReEDS regions, it is natural to ask whether these ITLs may simply be summed to obtain ITLs between larger regions, or whether the ITLs should be recalculated directly including all interface-crossing lines for the larger regions.

\begin{figure}
    \centering
    \includegraphics[width=1.0\linewidth]{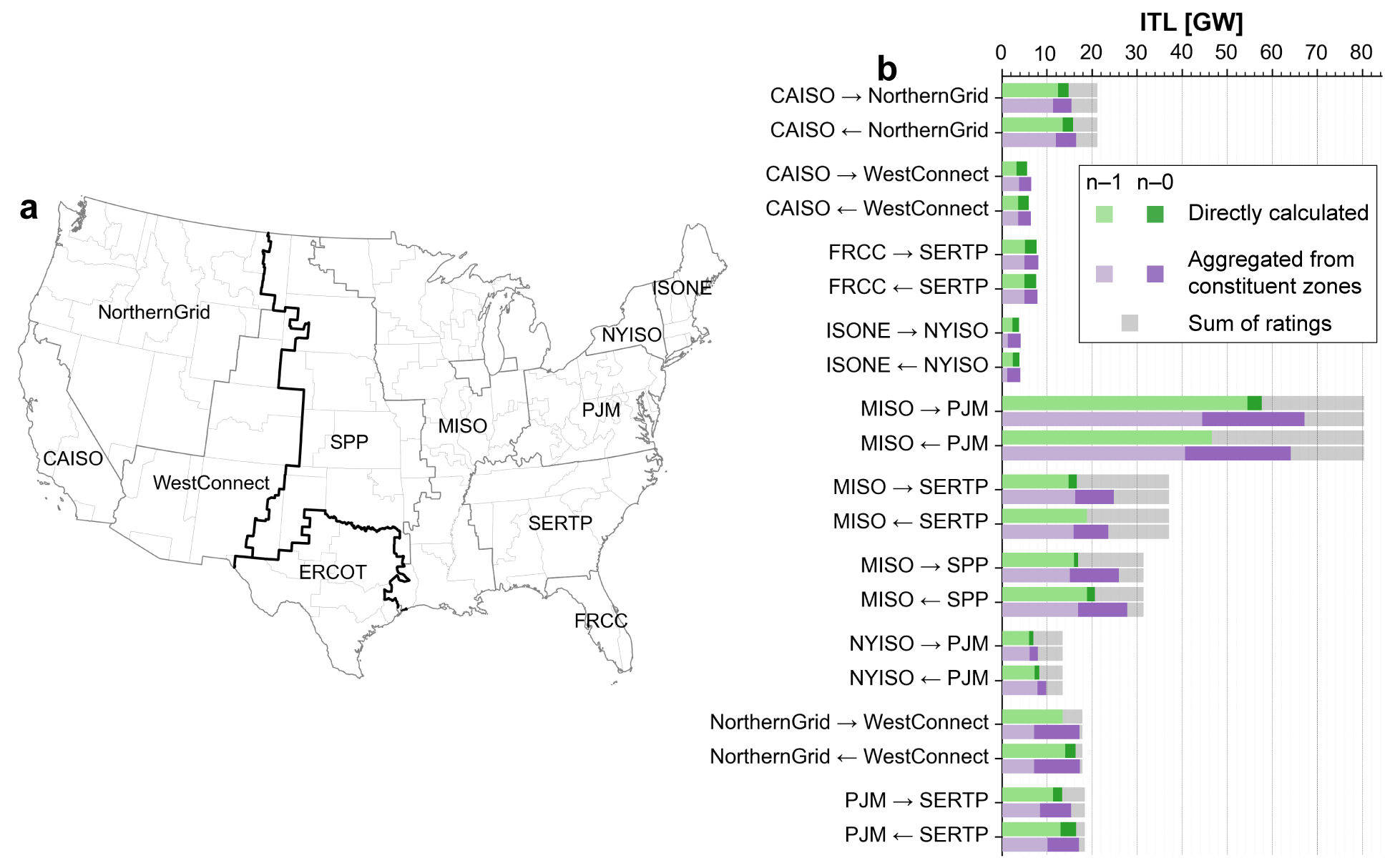}
    \caption{
        \textbf{Dependence of planning-region ITLs on calculation method.}
        \textbf{a}, Map of 11 planning regions, approximating FERC Order 1000 planning regions \cite{FERC1000} and ERCOT.
        Black lines indicate the boundaries between synchronous interconnections.
        \textbf{b}, ITLs between planning regions as a function of calculation method, direction, and contingency level.
        Green values were  directly calculated from flows on lines that cross the interfaces between planning regions;
        purple values were obtained by summing the ITLs of the underlying ReEDS zones;
        gray bars indicate the sum of ratings of interface-crossing lines.
        $n-1$ capacities are shown by lighter bars; additional capacity under $n-0$ conditions is shown by darker bars.
    }
    \label{fig:f_aggbar}
\end{figure}

Here we consider 11 ``planning regions'' that approximate the FERC Order 1000 planning regions \cite{FERC1000} and ERCOT.
These planning regions consist of groupings of the 134 ReEDS model zones (\textbf{\cref{fig:f_aggbar}a}).
\Cref{fig:f_aggbar}b shows the ITLs between these planning regions calculated using two different methods.
Values indicated by green bars were calculated directly from the flows on all lines crossing the interface between a given pair of planning regions.
Values indicated by purple bars are sums over the relevant ITLs between ReEDS zones shown in \cref{fig:f_map-zones_reeds,fig:f_mapdist-itl_reeds-contingency}; for example, the ``ISONE$\rightarrow$NYISO'' ITL under $n-1$ conditions is the sum of the p129$\rightarrow$p127, p131$\rightarrow$p127, and p132$\rightarrow$p127 ITLs under $n-1$ conditions.

Two consistent trends are apparent in \cref{fig:f_aggbar}b.
First, $n-0$ ITLs are uniformly higher when calculated by aggregating the ITLs of constituent zones than when calculated directly.
This trend is directly analogous to the trend observed in \cref{fig:f_mapdist-itl_reeds-contingency}c;
just as the sum of ratings of interface-crossing lines tends to overestimate the ITL,
the sum of ITLs of constituent regions tends to overestimate the ITL between larger regions.
The difference is larger for planning region interfaces that encompass more constituent-zone interfaces;
for example, the interfaces with MISO have a larger discrepancy between the two methods than the interfaces with NYISO.

Second, the $n-1$ ITLs are consistently lower when obtained from the sum of $n-1$ ITLs for constituent zones than when calculated directly.
The MISO$||$SPP interface encompasses 18 zonal interfaces, so the sum of those 18 $n-1$ capacities corresponds to a situation where 18 interface-crossing lines are not contributing to the ITL.
When the MISO$||$SPP $n-1$ ITL is calculated directly, only a single interface-crossing line is dropped.
As such, if an $n-1$ ITL is desired between larger aggregated regions, it should be calculated directly (as represented by the green bars in \cref{fig:f_aggbar}) rather than as a sum of the $n-1$ ITLs of constituent interfaces (as represented by the purple bars in \cref{fig:f_aggbar}).

\section{Conclusions} \label{sec:conc}

In summary, we have demonstrated a scalable method for calculating zonal interface transfer limits from nodal network transmission data that is applicable to arbitrary zone shapes and is compatible with different contingency levels.
Results for the \about\num{65000}-bus NARIS network (\cref{fig:f_mapdist-itl_reeds-contingency}) illustrate that while summing the ratings of interface-crossing lines provides a reasonable estimate of $n-0$ ITLs for some interfaces, it overestimates the ITL for the majority of interfaces.
In general, accounting for the limitations imposed by line reactances and bus injection/withdrawal capabilities helps improve the accuracy of estimated interface capacities and reduces the divergence between zonal models and more detailed nodal models.

The method presented here could be extended in several ways.
Comparing estimated ITLs to published path or interface ratings
could assist in cleaning of the underlying network data and improve the quality of the estimated ITLs.
The dual values of constraint \cref{eq:rating} provide the marginal increase in ITL that would result from increasing the rating of a particular line;
if combined with estimates of upgrade costs (e.g. via reconductoring), one could construct an ``upgrade supply curve'' that identifies the most cost-effective interventions for achieving a desired increase in ITL.
Modifying network parameters via FACTS could relax constraint \cref{eq:ptdf}, and adding energy storage or distributed energy resources (DERs) at buses without existing generation assets could alleviate constraint \cref{eq:bus};
both strategies could accordingly increase ITLs.
Adopting dynamic line ratings (DLR) could also increase ITLs but would increase complexity, given the need to repeat the ITL calculation for different weather conditions.

The greatest limitation of this method is its reliance on detailed nodal network data, which typically are not available publicly.
Necessary data fields for the application of this method include line reactance, line power rating, bus location, and connectivity between lines and buses; bus type, if available, can improve the estimate by allowing the use of constraint \cref{eq:bus}.
Some open-source synthetic network datasets are available (e.g. \cite{TAMU,Xu2020});
comparing ITLs calculated from actual network data with those calculated from synthetic networks could provide a useful validation metric for improving the quality of synthetic datasets.

\section*{Acknowledgments}
The authors acknowledge Luke Lavin, Hamody Hindi, Josh Novacheck, Leonardo Rese, and Mike Meshek for helpful discussions and feedback.
This work was authored by the National Renewable Energy Laboratory, operated by the Alliance for Sustainable Energy, LLC, for the U.S. Department of Energy (DOE) under contract no. DE-AC36-08GO28308.
Funding was provided by the U.S. Department of Energy Office of Energy Efficiency and Renewable Energy Wind Energy Technologies Office.
The views expressed in the article do not necessarily represent the views of the DOE or the U.S. Government.

\renewcommand\baselinestretch{1.2}\small\
\bibliographystyle{Brown2019-bibstyle}
\bibliography{NREL.bib}
\renewcommand\baselinestretch{1.8}\normalsize\

\newpage
\appendix

\setcounter{figure}{0}
\renewcommand{\thefigure}{S\arabic{figure}}
\setcounter{table}{0}
\renewcommand{\thetable}{S\arabic{table}}
\setcounter{equation}{0}
\renewcommand{\theequation}{S\arabic{equation}}
\renewcommand{\thesection}{S\arabic{section}}

\section*{Supplemental Information} \label{sec:SI}
\begin{figure}[htbp]
    \centering
    \includegraphics[width=0.7\linewidth]{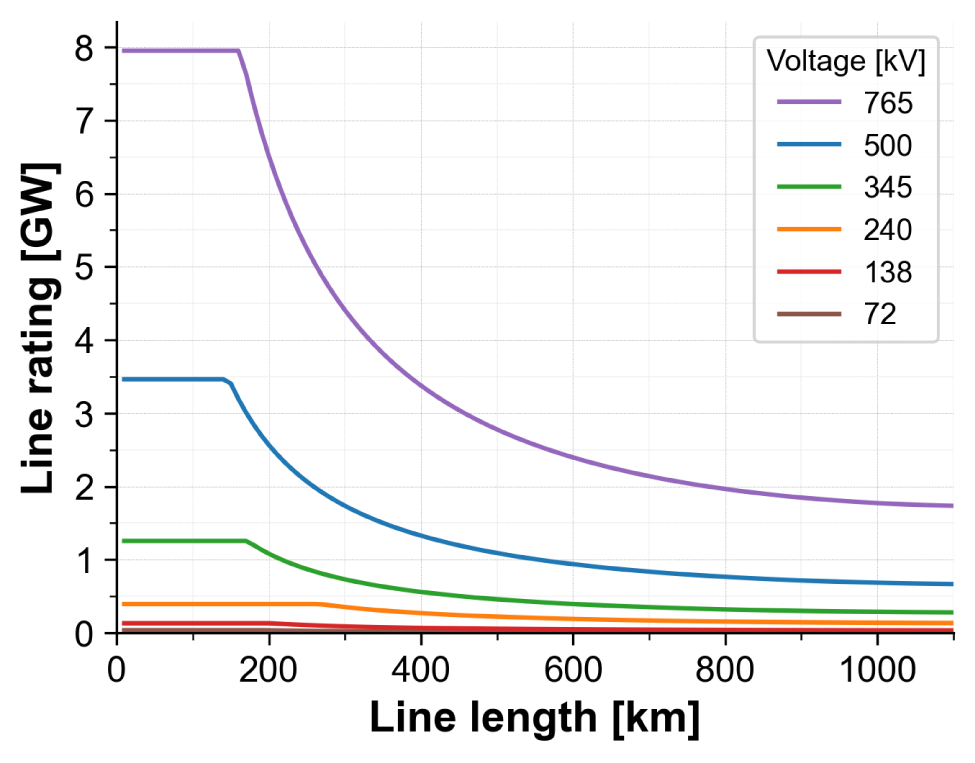}
    \caption{
        \textbf{Line loadability versus distance used to fill missing line ratings.}
    }
    \label{fig:f_loadability}
\end{figure}

\setstretch{1.2}
\textbf{\Cref{fig:f_loadability}} shows the line loadability curves used to fill missing line ratings as discussed in \cref{sec:data}.
Shorter lines are limited by their thermal rating (current carrying capability) while longer lines are limited by their stability limits.
For the purposes of this analysis, selected simplifications in the representation of lines were made to characterize these loadability curves and include:
\begin{itemize}
    \item Lossless lines (simplified representation of receiving-end to sending-end voltages as well as phase-angles). 
    \item A maximum voltage drop of \SI{5}{\percent} from receiving end to sending end.
    \item A maximum power angle of \SI{45}{\degree}~between the receiving end and sending end (to ensure a reasonable stability margin).  Although the absolute theoretical power angle limit under steady-state conditions is \SI{90}{\degree}, ranges of \SI{35}{\degree}--~\SI{45}{\degree}~are typical to ensure transient stability during large system disturbances.
    \item Voltage specific lumped model parameters from \cite{Hao2008}.
    \item The exclusion of voltage stability limits (only angular stability limits are included).
\end{itemize}


\end{document}